\documentclass[superscriptaddress,aps,journal=prl,preprint]{revtex4-2}
\usepackage{graphicx}
\usepackage{gensymb}
\usepackage{bm,amsmath,amssymb} 
\usepackage{physics}
\usepackage{color, soul} 
\usepackage{dcolumn}   
\usepackage{multirow}
\usepackage[colorlinks=true, linkcolor=blue, citecolor=blue, urlcolor=blue]{hyperref}

\newcommand{\etal}{\textit{et al.}}
\newcommand{\eg}{\textit{e.g.}}

\begin{document}

\title{Proximity Ferroelectricity Driven by Mobile High-Miller-Index Domain Walls}

\author{Changming Ke}
\affiliation{Department of Physics, School of Science, Westlake University, Hangzhou 310030, China}
\affiliation{Institute of Natural Sciences, Westlake Institute for Advanced Study, Hangzhou 310024, China}

\author{Shi Liu}
\email{liushi@westlake.edu.cn}
\affiliation{Department of Physics, School of Science, Westlake University, Hangzhou 310030, China}
\affiliation{Institute of Natural Sciences, Westlake Institute for Advanced Study, Hangzhou 310024, China}

\date{\today}

\begin{abstract}{
Wurtzite ferroelectrics such as scandium-doped aluminum nitride (AlScN) are promising for next-generation memory because of their compatibility with semiconductor processes and strong spontaneous polarization. Ferroelectric switching in these materials is typically attributed to doping-induced softening of the bulk switching barrier. However, recent reports of proximity ferroelectricity, in which undoped AlN layers up to 500 nm thick fully switch in AlN/AlScN multilayers, challenge this view. Here, we reveal an alternative switching mechanism mediated by high-Miller-index domain walls, long overlooked due to their complex geometry and presumed instability. Using first-principles calculations and machine-learning molecular dynamics simulations, we show that these walls, once nucleated, migrate with exceptionally low barriers. The Sc dopants play a dual role: they stabilize high-index walls and thereby promote nucleation, while also introducing pinning that hinders wall motion. In multilayers, our simulations demonstrate that mobile domain walls nucleated in AlScN can propagate deep into adjacent AlN, where they move easily without dopant pinning, enabling low-field switching across thick undoped layers. This microscopic divide-and-
conquer mechanism resolves the puzzle of proximity ferroelectricity and highlights high-index interfaces as an underexplored lever for controlling ferroelectric switching.
}
\end{abstract}
\maketitle
\newpage

Wurtzite aluminum nitride (AlN) has long been regarded as the archetype of a polar but non-ferroelectric material~\cite{Fichtner25p021310, FabiopR10024}, due to its  prohibitively high barrier to polarization reversal ~\cite{Liu23p122901, Cheng-wei24peadl0848, Ke23p616}. This conventional understanding was fundamentally challenged by the discovery of ferroelectricity in scandium-doped AlN (AlScN)~\cite{Fichtner19p114103}, which exhibits large remanent polarization ($>100$~$\mu$C/cm$^2$)~\cite{Fichtner19p114103, Mizutani21p105501}, a high Curie temperature ($T_c>1000$ K)~\cite{Islam21p232905}, and compatibility with standard semiconductor manufacturing~\cite{Kim23p422}. The realization of switchable polarization in AlScN has since motivated extensive efforts to identify other doped wurtzite ferroelectrics, including AlBN~\cite{Hayden21p044412, Wang21p111902, Calderon23p1034}, GaScN~\cite{Uehara21p172901, Wang21p111902}, AlYN~\cite{Wang23p033504}, ZnMgO~\cite{Ferri21p044101}, and AlGdN~\cite{Lee25p021114}. 

More recently, a paradigm known as ``proximity ferroelectricity" has emerged as a strategy to induce polarization reversal in pristine, nominally non-ferroelectric materials such as AlN and ZnO. In this approach, these strongly polar materials are incorporated into heterostructures with switchable layers, typically doped ferroelectric counterparts, that enable polarization switching~\cite{Skidmore25p574, Kim25pe09088}. 
Remarkably, experimental studies have demonstrated that an AlScN ferroelectric ``trigger'' layer constituting as little as 3\% of the total heterostructure volume can induce polarization reversal in an adjacent AlN layer up to 500~nm thick. Eliseev \etal~developed a thermodynamic framework that attributes proximity ferroelectricity to electrostatic coupling between the non-switchable polar layers and the adjacent doped ferroelectric component~\cite{Eliseev25p021058, Eliseev25p054026}. This coupling can destabilize the initially stable polarization state of the non-ferroelectric layers, thereby enabling collective switching across the multilayer structure.
While this continuum description establishes the thermodynamic feasibility of proximity-induced switching, it tends to overestimate the intrinsic coercive fields required for polarization reversal, often predicting values that exceed both experimental measurements and the dielectric-breakdown limit. Moreover, the model lacks atomistic resolution and does not account for the critical role of domain walls (DWs), whose nucleation and motion are increasingly recognized as the dominant kinetic mechanisms of polarization reversal in wurtzite ferroelectrics~\cite{Wang25p76, Lu25p2503143, Huang26p026801}.

By combining finite-field \textit{ab initio} molecular dynamics (AIMD) and large-scale deep potential molecular dynamics (DPMD) simulations, we reveal a microscopic divide-and-conquer mechanism for proximity ferroelectricity. We find that weakly charged DWs on high-Miller-index planes, such as \{11$\bar{2}$1\} and \{66$\bar{3}$2\}, refereed to here as high-index walls, are highly mobile and can drive polarization switching in pristine AlN at fields below 1.6~MV/cm. The key obstacle in pure AlN is therefore not wall motion, but wall formation: these DWs have prohibitively high formation energies and are thus difficult to nucleate even under strong electric fields. Sc doping solves this nucleation problem by substantially lowering the formation energy and stabilizing the high-index walls. At the same time, however, Sc dopants act as pinning centers that impede wall motion in the doped region.
Proximity ferroelectricity resolves this apparent contradiction through a divide-and-conquer strategy: the doped layer facilitates DW nucleation, while the adjacent undoped AlN layer provides a low-pinning medium in which the injected walls can propagate rapidly and reverse polarization. This mechanism is fundamentally different from the conventional picture in which Sc acts only by softening the double-well potential for homogeneous switching. We further predict analogous switching mediated by high-index DWs in wurtzite GaN and ZnO, with theoretical threshold fields down to $\approx$0.8~MV/cm and $\approx$1.1~MV/cm, respectively. Our results identify interface-assisted DW nucleation and injection as a general design principle for inducing ferroelectric switching in otherwise nonswitchable polar materials.

\begin{figure}[htb]
\centering
\includegraphics[scale=0.6, trim = 0 140 0 10, clip]{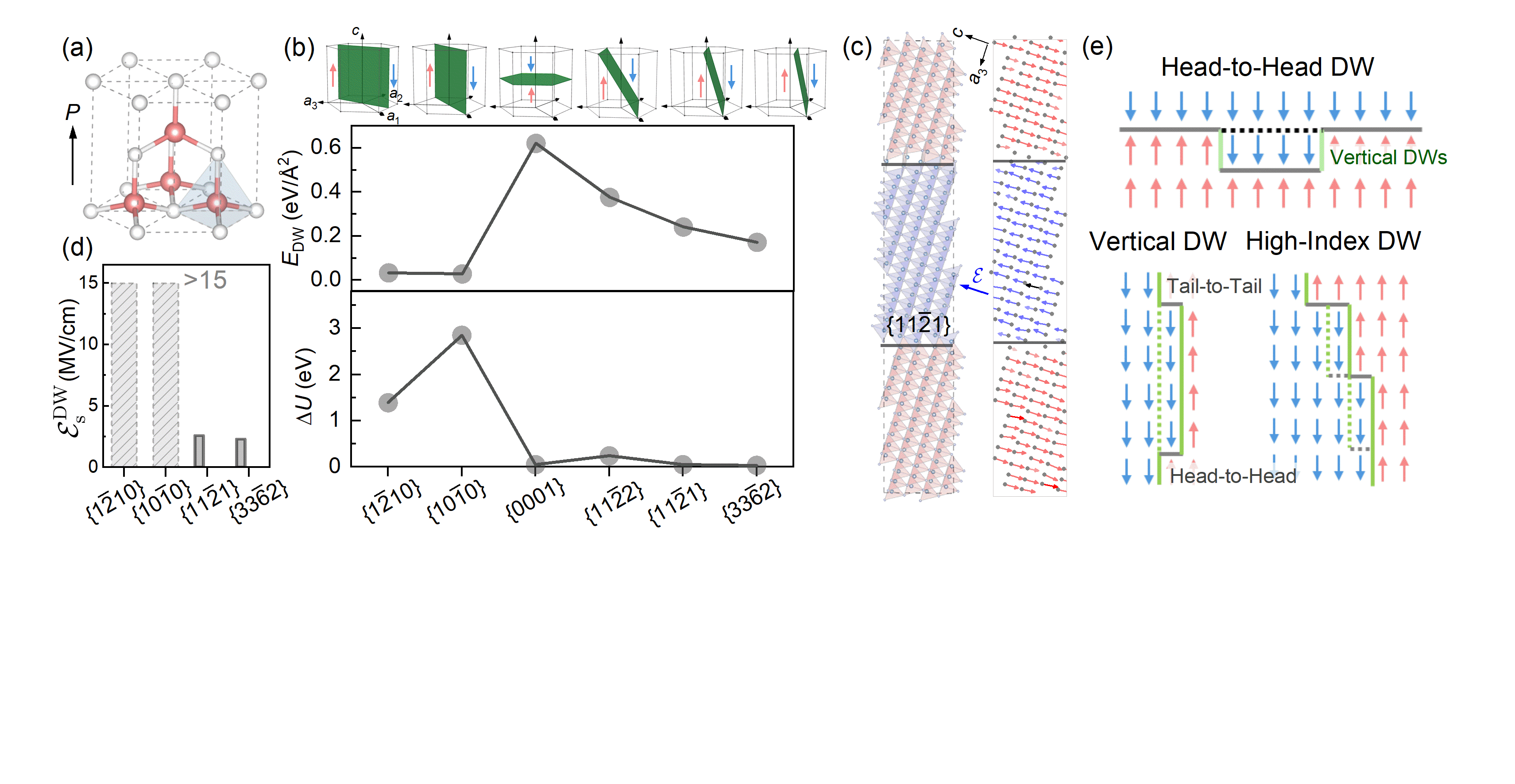}
 \caption{\textbf{High-Miller-index domain walls enable switching in pristine AlN at low electric fields.}
(a) Atomic structure of wurtzite AlN in the hexagonal lattice. Displacement of Al cations (red spheres) relative to the planes of N anions (white spheres) along the $c$ axis gives rise to an upward polarization $P$.
(b) Domain-wall (DW) energies ($E_{\mathrm{DW}}$) and corresponding migration barriers ($\Delta U$) for 180$^\circ$ walls, illustrating their overall negative correlation. The top panel shows the wall orientations in the hexagonal lattice. The $\{1\bar{2}10\}$ and $\{10\bar{1}0\}$ walls are conventional charge-neutral vertical 180$^\circ$ walls that separate domains with upward and downward polarization. By contrast, the head-to-head $\{0001\}$ wall is strongly charged and exhibits a prohibitively high energy because of electrostatic repulsion. DWs on high-Miller-index planes, such as $\{11\bar{2}2\}$, $\{11\bar{2}1\}$, and $\{33\bar{6}2\}$, are only weakly charged because only a small component of the polarization in neighboring domains is arranged in a head-to-head configuration.
(c) Atomic structure and polarization profile of a supercell containing a pair of $\{11\bar{2}1\}$ DWs used in finite-field \textit{ab initio} molecular dynamics (AIMD) simulations.
(d) Comparison of DW switching fields ($\mathcal{E}_{s}^{\rm DW}$) determined from AIMD simulations. The charge-neutral $\{1\bar{2}10\}$ and $\{10\bar{1}0\}$ walls do not move within 5~ps even under fields as high as 15~MV/cm, whereas the high-index $\{11\bar{2}2\}$ and $\{33\bar{6}2\}$ walls are mobile at a much lower field of 2.5~MV/cm.
(e) Schematic illustration of the interfacial changes associated with the motion of charge-neutral vertical, head-to-head, and high-index DWs. The diffuse polarization profile across each wall is omitted for clarity. Nucleation on a low-energy vertical DW requires the formation of high-energy head-to-head or tail-to-tail walls, whereas motion of a high-energy head-to-head DW creates only low-energy vertical walls. By contrast, motion of a high-index DW does not generate additional DWs within this simplified picture.}
\label{fig_AlN}
\end{figure}

We begin by considering the DW geometry in wurtzite AlN, which crystallizes in a hexagonal unit cell where the cation sublattice is displaced relative to the anion planes along the polar $c$-axis, as illustrated in Fig.\ref{fig_AlN}(a). We first focus on three principal crystallographic planes: the non-polar prismatic planes $\{1\bar{2}10\}$ and $\{10\bar{1}0\}$, and the polar basal plane $\{0001\}$ [see the top panel of Fig.\ref{fig_AlN}(b)]. These planes define three distinct types of 180\degree~DWs separating domains with antiparallel polarization vectors. Using density functional theory (DFT; see details in in Supplementary Material~\cite{SIforL62}), we computed the formation energies ($E_{\rm DW}$) and migration barriers ($\Delta U$) of these DWs. As shown in Fig.~\ref{fig_AlN}(b), the charge-neutral vertical walls ($\{1\bar{2}10\}$ and $\{10\bar{1}0\}$) exhibit low formation energies but are kinetically inactive due to prohibitively high migration barriers ($>1$~eV). In contrast, the charged head-to-head $\{0001\}$ wall has a much higher formation energy due to strong electrostatic repulsion at the wall, but displays a negligible migration barrier. 

Planes with high Miller indices, such as $\{11\bar{2}1\}$ and $\{33\bar{6}2\}$, are only slightly tilted relative to the neighboring polarization vectors and therefore form so-called weakly charged DWs, across which only a small component of the polarization adopts a head-to-head (or tail-to-tail) configuration. These high-index DWs still satisfy the 180\degree~condition and separate antiparallel domains, as illustrated schematically in Fig.~\ref{fig_AlN}(b). Traditionally, such high-index DWs have received little attention in ferroelectrics. However, our DFT calculations show that high-index DWs in AlN have substantially lower formation energies ($\approx 0.15$~eV/\AA$^2$) than the $\{0001\}$ wall while retaining low migration barriers (below 0.1~eV), significantly smaller than those of the charge-neutral $\{1\bar{2}10\}$ and $\{10\bar{1}0\}$ walls.

To further assess the mobility of the different 180\degree~DWs, we perform finite-field AIMD simulations using the force method~\cite{Umari02p157602}, as implemented in the Vienna ab initio Simulation Package~\cite{Kresse96p11169, Kresse96p15}.
Specifically, the electric-field-induced force is treated as an external force added to each atom based on the product of the field strength and Born effective charge tensors (see additional details in~\cite{SIforL62}).  
The supercell containing high-index $\{11\bar{2}1\}$ walls is shown in Fig.~\ref{fig_AlN}(c), and the electric field $\mathcal{E}$ is applied long the polar axis. 
Because AIMD is computationally demanding, we use the switching field, $\mathcal{E}_s^{\rm DW}$, defined as the minimum electric field required to induce wall motion within 5 ps, as a first-principles metric for DW mobility. This quantity is not intended to represent an experimental coercive field, but to provide a practical measure of how readily a given DW can move on AIMD-accessible time scales. We expect $\mathcal{E}_s^{\rm DW}$ to be much higher than coercive field as the electric field is commonly applied for several microseconds in experiments. 
As shown in Fig.~\ref{fig_AlN}(d), the charge-neutral DWs remain immobile under fields up to 15~MV/cm. By contrast, the high-index $\{11\bar{2}1\}$ and $\{33\bar{6}2\}$ DWs respond at fields below 2.5~MV/cm. These results demonstrate that, once stabilized, high-index DWs in AlN are highly mobile and can enable polarization switching at substantially reduced fields.

The present results also clarify how DW stability is related to mobility. In ferroelectrics, our understanding of DW kinetics has been shaped largely by studies of perovskite oxides (\eg, BaTiO$_3$ and PbTiO$_3$), where lower wall energy is often associated with higher mobility~\cite{Liu16p360}. However, such comparisons usually involve fundamentally different wall types, such as 90$^\circ$ and 180$^\circ$ DWs, with the former often having both lower formation energy and higher mobility. By contrast, our results for 180$^\circ$ DWs in AlN reveal the opposite trend: more stable walls are generally less mobile. This behavior is in fact more consistent with a simple energetic picture: a more stable wall should be intrinsically more resistant to external perturbations and therefore more difficult to displace.
It is well established that DW migration is governed by the local nucleation process during wall translation, namely, by the energy cost of the newly formed interface~\cite{Shin18p075001,Liu16p360}. Figure~\ref{fig_AlN}(e) schematically compares the interfacial changes involved in the motion of charge-neutral vertical, head-to-head, and high-index DWs, while neglecting the diffuseness of the polarization profile across the wall for simplicity. For a vertical DW to move, segments of high-energy head-to-head and tail-to-tail walls (gray lines) must be created, which results in a large migration barrier. In contrast, motion of the charged head-to-head DW proceeds through the formation of low-energy vertical wall segments (green lines) and therefore encounters only a small barrier. Interestingly, in the simplest case, motion of a high-index DW involves little change in interfacial energy, which naturally accounts for its high mobility.

\begin{figure}[htb]
\centering
\includegraphics[scale=1, trim = 0 30 0 0, clip]{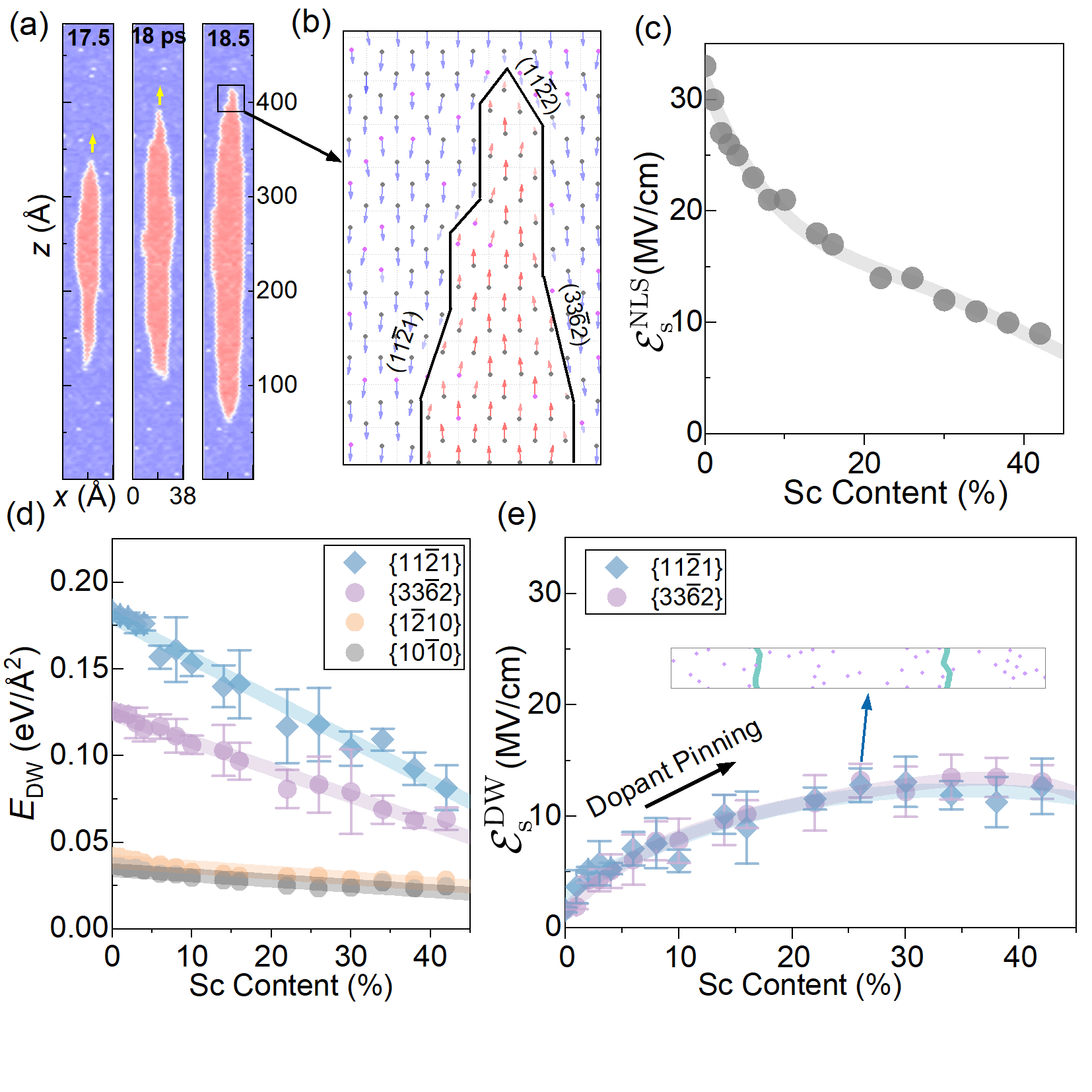}
 \caption{\textbf{Dual role of Sc doping in domain nucleation and DW motion.}
(a) Domain nucleation in DPMD simulations of Al$_{0.78}$Sc$_{0.22}$N. The initial state is a single domain with polarization pointing along $-z$. Under an upward electric field, a needle-like reversed nucleus (red) forms within the parent domain (blue), and its growth is dominated by the motion of the needle tip along the $c$ axis.
(b) Spatial polarization profiles show that the DWs bounding the nucleus exhibit a jagged morphology composed of high-index $\{11\bar{2}1\}$, $\{11\bar{2}2\}$, and $\{33\bar{6}2\}$ facets.
(c) Simulated nucleation-limited switching field, $\mathcal{E}_s^{\mathrm{NLS}}$ (see definition in the main text), decreases monotonically with increasing Sc content.
(d) Domain-wall energy, $E_{\mathrm{DW}}$, as a function of Sc concentration. The energies of all walls decrease with increasing Sc content, indicating that Sc doping can lower the nucleation barrier.
(e) Pinning effect of Sc dopants on high-index $\{11\bar{2}1\}$ and $\{33\bar{6}2\}$ walls. The electric field required to drive DW motion, $\mathcal{E}_s^{\mathrm{DW}}$, increases with Sc content, indicating that Sc dopants act as pinning centers that hinder DW motion (inset). Remarkably, in pure AlN, where dopant pinning is absent, these high-index DWs can propagate under a much lower electric field of 1.6~MV/cm, provided that they are present.
}
  \label{fig_nucleus}
\end{figure}

The inverse correlation between 180$^\circ$ DW stability and mobility has important implications for understanding the role of Sc doping in AlScN. Specifically, Sc doping may facilitate domain nucleation by lowering the DW energy, while simultaneously hindering DW motion by stabilizing the walls and thereby giving rise to a pinning effect. To investigate these competing effects, we developed a deep potential (DP) model that accurately captures interatomic interactions in AlScN over a broad range of Sc concentrations, reproducing domain-wall formation energies and migration barriers in close agreement with DFT calculations (see force-field validation tests in Supplementary Material~\cite{SIforL62}). We then performed DPMD simulations of polarization reversal in an $8\times8\times72$ supercell containing 55,296 atoms of Al$_{0.78}$Sc$_{0.22}$N under an electric field applied along the polar $c$-axis. As shown in Fig.~\ref{fig_nucleus}(a), polarization reversal proceeds primarily through the nucleation of a needle-like reversed domain. Atomistic analysis [Fig.~\ref{fig_nucleus}(b)] further reveals that the tip of the nucleus preferentially adopts jagged high-Miller-index facets corresponding to $\{11\bar{2}1\}$, $\{11\bar{2}2\}$, and $\{33\bar{6}2\}$ domain walls. Importantly, the longitudinal growth of the nucleus, driven by the propagation of these high-index walls, is much faster than its lateral expansion, which requires the motion of charge-neutral vertical DWs. This is consistent with the trend in the DW migration barriers obtained from the DFT calculations.

To quantify the effect of Sc doping on domain nucleation, we define the nucleation-limited switching field, $\mathcal{E}_s^{\mathrm{NLS}}$, as the minimum electric field required to initiate domain nucleation and sustain its growth along the polar axis within a fixed simulation time of 100~ps. As shown in Fig.~\ref{fig_nucleus}(c), $\mathcal{E}_s^{\mathrm{NLS}}$ decreases monotonically with increasing Sc concentration, consistent with experimental observations that the coercive field decreases with increasing Sc content. This reduction in $\mathcal{E}_s^{\mathrm{NLS}}$ is mainly attributed to the doping-induced lowering of the formation energies of the high-index 180$^\circ$ DWs, such as $\{11\bar{2}1\}$ and $\{33\bar{6}2\}$, which form the tip of the nucleus and dominate the nucleation barrier. In pristine AlN, these walls have such high formation energies that a field exceeding 30~MV/cm is required in DPMD simulations to nucleate them [Fig.~\ref{fig_nucleus}(c)]. This interpretation is supported by the calculated formation energies shown in Fig.~\ref{fig_nucleus}(d), where Sc doping substantially lowers the formation energies of the high-index walls. These results indicate that Sc doping promotes ferroelectric switching primarily by facilitating domain nucleation.

Given the inverse relationship between DW stability and mobility established by our DFT calculations, Sc doping is expected to impede the motion of high-index DWs. This prediction is further confirmed by our DPMD simulations using supercells containing pre-existing $\{11\bar{2}1\}$ or $\{33\bar{6}2\}$ walls. As shown in Fig.~\ref{fig_nucleus}(e), the switching field required to activate the motion of these walls, $\mathcal{E}_s^{\mathrm{DW}}$, defined as the minimum field needed to initiate DW motion within 100~ps, increases with Sc concentration.
These results reveal a dual role of Sc in AlScN. Thermodynamically, Sc promotes switching by stabilizing high-index DWs and thereby lowering the nucleation barrier, which reduces $\mathcal{E}_s^{\mathrm{NLS}}$. Kinetically, it suppresses DW motion through enhanced pinning, making nucleus growth more difficult. Nevertheless, we suggest that in AlScN, the switching process is primarily limited by domain nucleation, as indicated by $\mathcal{E}_s^{\rm NLS} > \mathcal{E}_s^{\rm DW}$ at low Sc concentrations. Therefore, the experimentally observed reduction in coercive field with increasing Sc doping is governed mainly by the lowered nucleation barrier, while the reduced mobility of high-index walls plays a less significant role.

We note that experimental studies of switching dynamics in wurtzite ferroelectrics have reported anomalously large Avrami exponents for polarization reversal (\eg, $n=11$), far exceeding the ranges predicted by conventional KAI (Kolmogorov--Avrami--Ishibashi, $1<n<4$) or NLS ($n<1$) models~\cite{Yazawa23p2936}. Such unusually high exponents have been attributed to nucleation and domain growth occurring on comparable timescales. Our DPMD simulations provide a natural explanation for this behavior. The comparison of Figs.~\ref{fig_nucleus}(c) and \ref{fig_nucleus}(e) shows that when the Sc concentration exceeds 25\%, the characteristic switching fields for nucleation and domain-wall motion become comparable, $\mathcal{E}_s^{\rm NLS} \approx \mathcal{E}_s^{\rm DW}$, so that neither process clearly dominates the overall switching dynamics.

The dual role of Sc doping in domain nucleation and DW motion provides a useful framework for understanding proximity ferroelectricity in AlN/AlScN multilayers. We propose that this behavior follows a divide-and-conquer mechanism: the doped AlScN region enables easy nucleation, while the undoped AlN region supports rapid DW propagation. Specifically, the low nucleation barrier in AlScN facilitates the formation of reversed domains; as these nuclei elongate along the polar axis, they inject high-index DWs into the adjacent pristine AlN layers, where the walls can propagate rapidly because of their intrinsically high mobility and the absence of Sc-induced pinning.

\begin{figure}[htb]
\centering
\includegraphics[scale=0.5, trim = 0 0 30 5, clip]{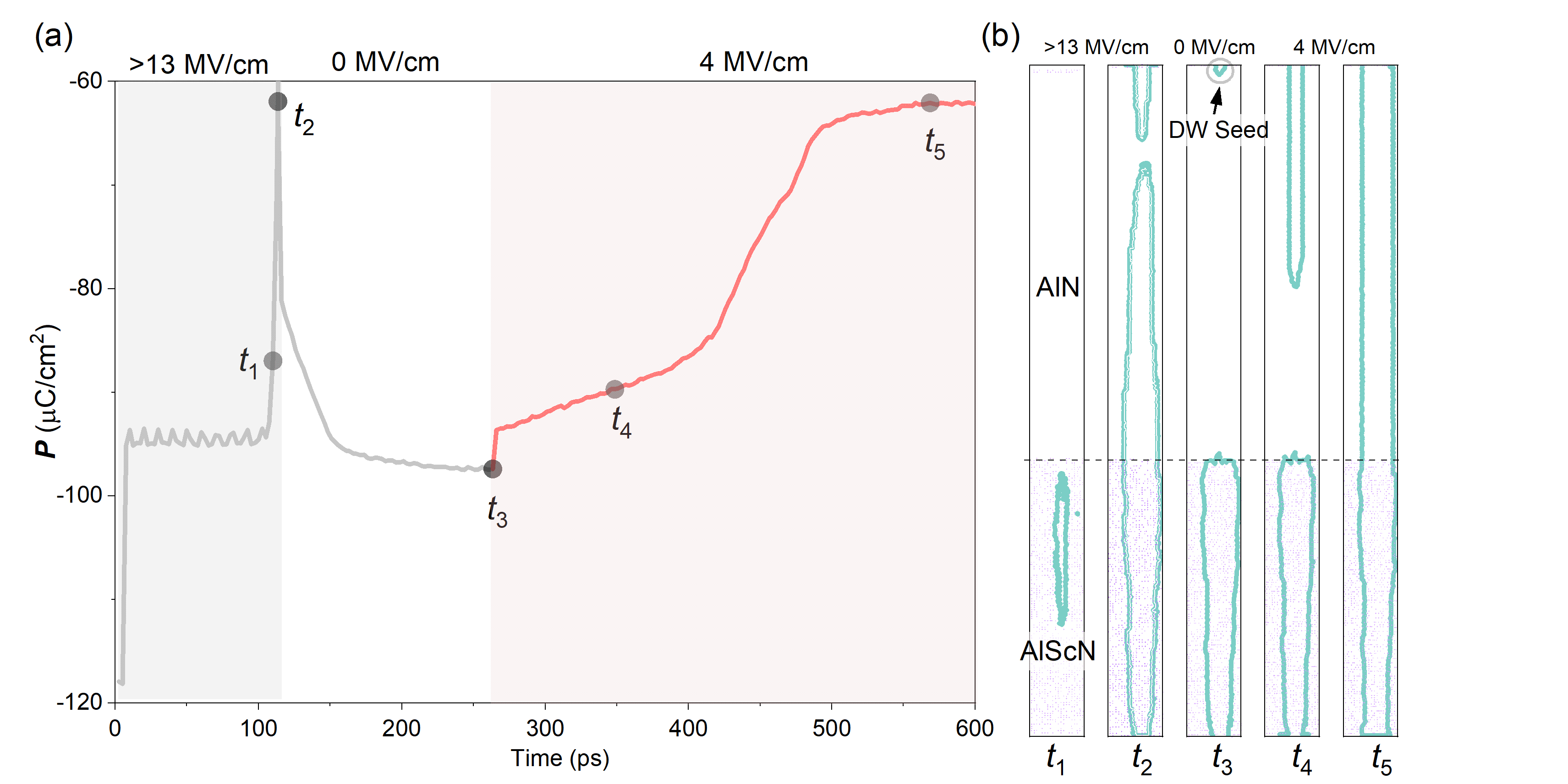}
  \caption{\textbf{Proximity ferroelectricity in AlN/Al$_{0.6}$Sc$_{0.4}$N originates from the injection of high-index DWs.}
(a) Polarization response of an AlN/Al$_{0.6}$Sc$_{0.4}$N heterostructure, consisting of a 100-unit-cell-thick AlN layer and a 60-unit-cell-thick Al$_{0.6}$Sc$_{0.4}$N layer, under a two-step electric-field pulse sequence in DPMD simulations.
(b) Snapshots of the heterostructure during the switching process. In the first step, a high electric field is applied to initiate polarization switching within 100~ps. Nucleation occurs in the Al$_{0.6}$Sc$_{0.4}$N layer at $t_1$. The tip of the nucleus then penetrates into the pristine AlN layer, where it drives polarization reversal. The field is intentionally turned off at $t_2$. Although the switched region rapidly shrinks, a metastable high-index DW seed remains at the interface. In the second step, a much lower electric field of 4~MV/cm is sufficient to drive DW motion in the pristine AlN layer. }
   \label{fig_proxi}
 \end{figure}

Our DPMD simulations of an AlN/Al$_{0.6}$Sc$_{0.4}$N heterostructure containing 256,000 atoms confirm this mechanism using a two-step electric-field pulse sequence, as illustrated in Fig.~\ref{fig_proxi}(a), with corresponding snapshots shown in Fig.~\ref{fig_proxi}(b).
Under an initial high-field pulse of 16~MV/cm, a needle-like reversed domain nucleates rapidly within the Al$_{0.6}$Sc$_{0.4}$N layer. Driven by the applied field, the nucleus grows vertically through the propagation of high-index DWs and eventually penetrates into the pristine AlN region at $t_1$. We intentionally turn off the field at $t_2$, before switching in the AlN layer is complete, to test whether the injected DWs can survive without external bias and subsequently reactivate switching. Although the switched region shrinks significantly after field removal, a metastable high-index DW seed remains at the AlN//Al$_{0.6}$Sc$_{0.4}$N interface. Remarkably, when a lower field of 4~MV/cm is applied at $t_3$, polarization reversal in the pristine AlN resumes through the propagation of these injected high-index DWs, as shown by the snapshots at $t_4$ and $t_5$. These results demonstrate that the doped layer acts as a DW injector by nucleating reversed domains efficiently and transferring mobile high-index DWs into the neighboring AlN layer to drive polarization reversal.

\begin{figure}[htb]
\centering
\includegraphics[scale=0.5, trim = 10 400 0 20, clip]{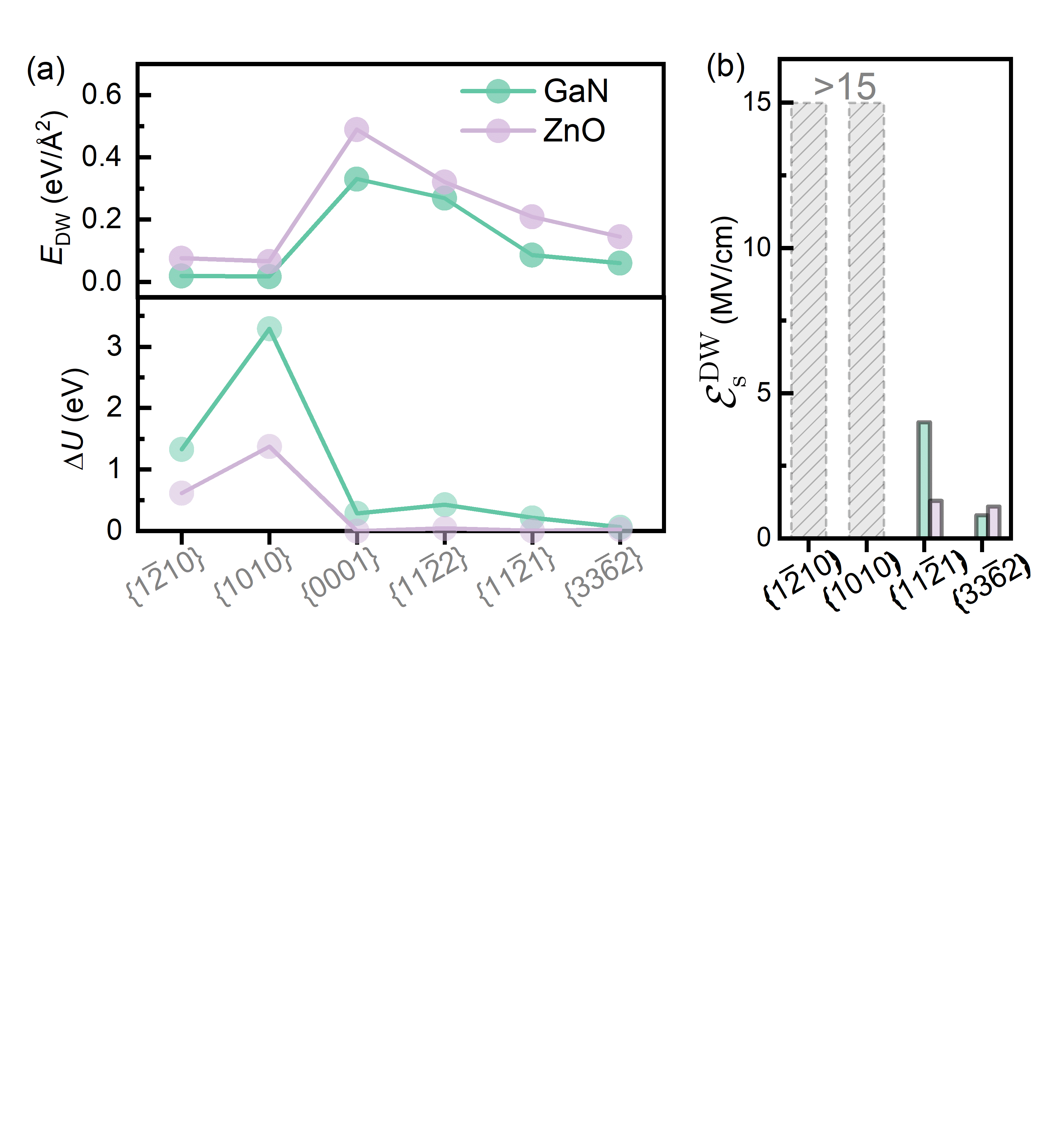}
 \caption{\textbf{Mobile high-index DWs in polar wurtzites.}
(a) DFT-calculated DW energies and migration barriers for GaN and ZnO. The trends are similar to those in AlN: high-index DWs, such as $\{11\bar{2}1\}$ and $\{33\bar{6}2\}$, exhibit much lower migration barriers than conventional charge-neutral vertical walls.
(b) Switching fields predicted from finite-field AIMD simulations for GaN and ZnO. High-index DWs exhibit substantially lower switching fields than charge-neutral vertical walls.
}
\label{fig_pres}
\end{figure}

Finally, we propose that this divide-and-conquer mechanism is a general feature of wurtzite-structured binary compounds. As shown in Fig.~\ref{fig_pres}(a), our DFT calculations for GaN and ZnO reveal an energetic landscape similar to that of AlN: high-index DWs such as $\{11\bar{2}1\}$ and $\{33\bar{6}2\}$ consistently exhibit moderate formation energies and low migration barriers. Using the same protocol as in Fig.~\ref{fig_AlN}(d), our finite-field AIMD simulations further confirm that, once such walls are nucleated, polarization switching in pristine GaN and ZnO can proceed at remarkably low fields of approximately 0.8~MV/cm and 1.1~MV/cm, respectively [Fig.~\ref{fig_pres}(b)].

In summary, we demonstrate an intrinsic negative correlation between the energy and mobility of 180$^\circ$ domain walls: walls with lower formation energies are generally more stable but more difficult to move, whereas higher-energy walls can exhibit substantially greater mobility. Although this trend is a natural consequence of the underlying energetics, it has long been obscured by comparisons between fundamentally different classes of DWs, such as 90$^\circ$ and 180$^\circ$ walls in PbTiO$_3$. In AlScN, this trade-off gives rise to a dual role of Sc doping. By lowering the energies of high-index domain walls, Sc thermodynamically facilitates the nucleation of reversed domains; at the same time, the increased wall stability suppresses their mobility through enhanced pinning. This intrinsic competition between domain nucleation and DW propagation provides an explanation for the anomalously high Avrami exponents observed in these systems.

More broadly, our results highlight that high-index DWs are not merely transitional structures in polarization reversal, but functional interfaces with distinct energetic and kinetic advantages. Their exceptionally high mobility make them especially effective for mediating fast switching once nucleated. From this perspective, proximity ferroelectricity can be understood as a DW-engineering strategy: doped layers act as nucleation sources that generate and inject mobile high-index walls into adjacent regions free of dopant-induced pinning, where they efficiently drive polarization reversal. Beyond offering a pathway to reduce coercive fields in wurtzite ferroelectrics, our work also identifies high-Miller-index interfaces as a promising and largely unexplored design space for controlling switching kinetics and realizing new ferroelectric functionalities.

\newpage
\begin{acknowledgments}
This work acknowledge the supports from National Natural Science Foundation of China (12574105) and Zhejiang Provincial Natural Science Foundation of China (LR25A040004). The computational resource is provided by Westlake HPC Center.
\end{acknowledgments}
\newpage

\bibliography{SL}

\newpage

\newpage

\clearpage
\newpage

\clearpage
\newpage

\end{document}